\documentclass[sigconf]{acmart}

\usepackage{booktabs} % For formal tables
\usepackage{flexisym}
\usepackage{bbm}
\usepackage{balance}
\usepackage{graphicx}
\usepackage{multirow}
\usepackage{float}
\usepackage{figsize}
\usepackage{amsmath}
\usepackage{todonotes}

\renewcommand\footnotetextcopyrightpermission[1]{}

\begin{document}
%\fancyhead{}
%\settopmatter{printacmref=true, printfolios=true}
\fancyhead{}

%\title{Improving Long-tail %Recommendation Over Time}
\title{The Unfairness of Popularity Bias in Recommendation}
% Reducing Popularity Bias: An Incremental Adaptation Approach
% If Not Today, Perhaps Tomorrow: Reducing Recommendation Popularity Bias Over Time
% Not Recommended Today, Come Back Tomorrow: Reducing Popularity Bias Over Time
% 
\titlenote{Copyright 2019 for this paper by its authors. Use permitted under Creative Commons License Attribution 4.0 International (CC BY 4.0).\\Presented at the RMSE workshop held in conjunction with the 13th ACM Conference on Recommender Systems (RecSys), 2019, in Copenhagen, Denmark.}
\author{Himan Abdollahpouri}
\affiliation{
\institution{University of Colorado Boulder}
\city{Boulder}
\state{USA}
}
\email{himan.abdollahpouri@colorado.edu}
 \author{Masoud Mansoury}

\affiliation{%
  \institution{Eindhoven University of Technology}
  \city{Eindhoven}
  \state{Netherlands}
}
\email{m.mansoury@tue.nl}
 \author{Robin Burke}
\affiliation{
\institution{University of Colorado Boulder}
\city{Boulder}
\state{USA}
}
\email{robin.burke@colorado.edu}
 \author{Bamshad Mobasher}
\affiliation{
\institution{DePaul University}
\city{Chicago}
  \state{USA}
}
\email{mobasher@cs.depaul.edu}
\begin{abstract}
Recommender systems are known to suffer from the popularity bias problem: popular (i.e. frequently rated) items get a lot of exposure while less popular ones are under-represented in the recommendations. Research in this area has been mainly focusing on finding ways to tackle this issue by increasing the number of recommended long-tail items or otherwise the overall catalog coverage. In this paper, however, we look at this problem from the users' perspective: we want to see how popularity bias causes the recommendations to deviate from what the user expects to get from the recommender system. We define three different groups of users according to their interest in popular items (Niche, Diverse and Blockbuster-focused) and show the impact of popularity bias on the users in each group. Our experimental results on a movie dataset show that in many recommendation algorithms the recommendations the users get are extremely concentrated on popular items even if a user is interested in long-tail and non-popular items showing an extreme bias disparity.     
\end{abstract}

\keywords{Recommender systems; Popularity bias; Fairness; Long-tail recommendation}

\maketitle
\section{Introduction}
Recommender systems have been widely used in a variety of different domains such as movies, music, online dating etc. Their goal is to help users find relevant items which are difficult or otherwise time-consuming to find in the absence of such systems.

Different types of algorithms are being used for recommendation depending on the domain or other constraints such as the availability of the data about users or items. One of the most widely-used classes of algorithms for recommendation is collaborative filtering. In these algorithms the recommendations for each user are being generated using the rating information from other users and items. Unlike other types of algorithms such as content-based recommendation, collaborative algorithms do not use the content information. 

One of limitations of collaborative recommenders is the problem of popularity bias \cite{longtailnichesriche,himanaies}: popular items are being recommended too frequently while the majority of other items do not get the deserved attention. 

Popularity bias could be problematic for a variety of different reasons: Long-tail (non-popular) items are important for generating a fuller understanding of users' preferences. Systems that use active learning to explore each user's profile will typically need to present more long tail items because these are the ones that the user is less likely to have rated, and where user's preferences are more likely to be diverse~\cite{nguyen2014exploring, resnick2013bursting}. 

In addition, long-tail recommendation can also be understood as a social good. A market that suffers from popularity bias will lack opportunities to discover more obscure products and will be, by definition, dominated by a few large brands or well-known artists~\cite{celma2008hits}. Such a market will be more homogeneous and offer fewer opportunities for innovation and creativity.

\begin{figure}
    \centering
    \includegraphics[height=1.1in, width=3.1in]{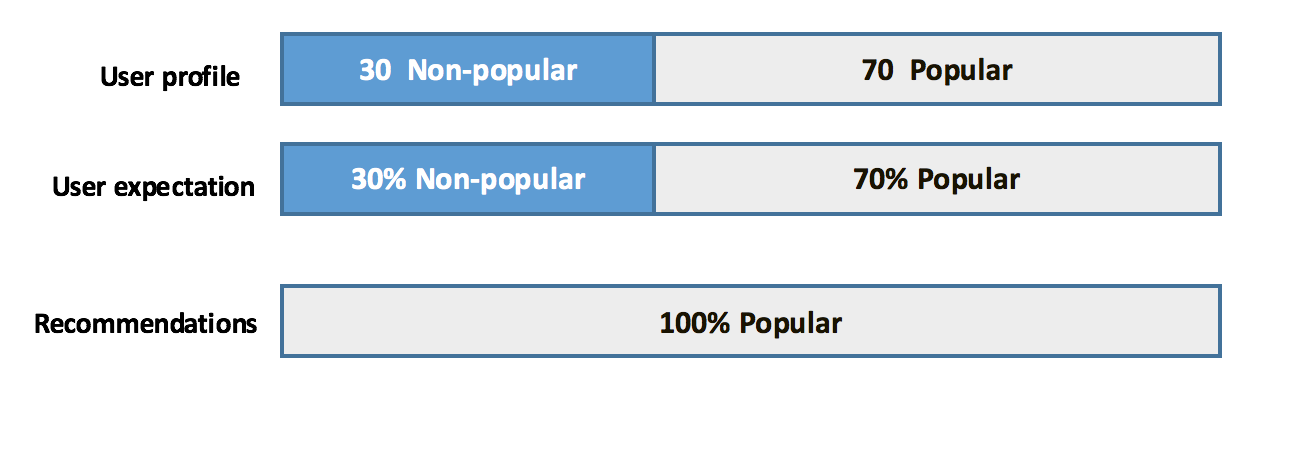}
    \caption{The inconsistency between user's expectation and the given recommendations}
    \label{calibration}
\end{figure}
In this paper, however, we look at the popularity bias from a different perspective: the users'. Look at figure ~\ref{calibration} for example: the user has rated 30 long-tail (non-popular) items and 70 popular items. So it is reasonable to expect the recommendations to keep the same ratio of popular and non-popular items. However, most of the recommendation algorithms produce a list of recommendations that is over-concentrated on popular items (often it is even close to 100\% popular items). In this paper, we show how different recommendation algorithms are propagating this bias differently for different users. In particular, we define three groups of users according to their degree of interest towards popular items and show the bias propagation for some groups is higher than the others. For instance, the niche users (users with the lowest degree of interest towards popular items) are affected the most by this bias. We also show these niche users and, in general users with lower interest in popular items, are more active (they have rated more items) in the system and therefore they should be considered as important stakeholders whose needs should be addressed properly by the recommender system. However, looking at several recommendation algorithms, we see these users are not getting the type of recommendations that they expect to get.
In summary, here are the important research questions we would like to answer in this paper:
\begin{itemize}
    \item \textbf{RQ1:} How much are different individuals or groups of users interested in popular items?
    \item \textbf{RQ2:} How do algorithms' popularity biases impact users with different degrees of interest in popular items?

\end{itemize}
To answer these questions we look at the movieLens 1M dataset and analyze the users' profiles in terms of their rating behavior. We also compare several recommendation algorithms with respect to their general popularity bias propagation and the extent to which this bias propagation is affecting different users. 

\section{Related Work}
The problem of popularity bias and the challenges it creates for the recommender system has been well studied by other researchers \cite{anderson2006long,brynjolfsson2006niches,longtailrecsys}. Authors in the mentioned works have mainly explored the overall accuracy of the recommendations in the presence of long-tail distribution in rating data. Moreover, some other researchers have proposed algorithms that can control this bias and give more chance to long-tail items to be recommended \cite{10.1109/TKDE.2011.15,DBLP:conf/recsys/KamishimaAAS14, abdollahpouri2017controlling,flairs2019}. 

In this work, however, we focus on the fairness of recommendations with respect to users' expectations. That is, we want to see how popularity bias in the input data is causing the recommendations to deviate from the actual interests of users with respect to how many popular items they expect to see in the recommended list. We call this deviation unfair because it is caused by the existing bias in data and also the algorithmic bias in some of the recommendation algorithms. A similar work to ours is \cite{steck2018calibrated} where author proposed the idea of calibration: the recommendations should be consistent with the spectrum of items a user has rated. For example, if a user has rated 70\% action movies and 30\% romance, it is expected to see the same pattern in the recommendations. Our work is different as we do not use content information to see how different algorithms are calibrated. In fact, Our work could be considered as an explanation for the problem discussed relative to genre calibration, namely that genre distortion in recommendations could be caused by different levels of popularity across genres.

Moreover, the concept of fairness in recommendation has been also gaining a lot of attention recently \cite{kamishima2012fairness,DBLP:journals/corr/YaoH17}. For example, finding solutions that removes algorithmic discrimination against users belong to a certain demographic information \cite{zhu2018fairness} or making sure items from different categories (e.g. long tail items or items belong to different providers)\cite{liu2018personalizing,burke2017balanced} are getting a fair exposure in the recommendations. Our definition of fairness in this paper is more related to the \textit{calibration fairness} \cite{steck2018calibrated} which means users should get the recommendations that are close to what they are expecting to get based on the type of items they have rated. 

And finally, Jennach et al. \cite{jannach2015recommenders} compared different recommendation algorithms in terms of accuracy and popularity bias. In that paper they observed some algorithms concentrate more on popular items than the others. In our work, we are mainly interested in seeing the popularity bias from the users' expectations perspective. 

\section{Popularity Bias in Data}
For all of the following sections we  use the MovieLens 1M dataset for our analysis which contains 1,000,209 anonymous ratings of approximately 3,900 movies made by 6,040 MovieLens users~\cite{movielens}.

\begin{figure}
    \centering
    \includegraphics[height=1.7in]{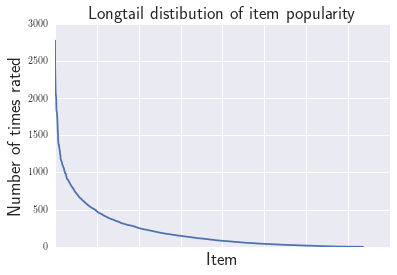}
    \caption{The long-tail of item popularity in rating data}
    \label{longtail}
\end{figure}

Rating data is generally skewed towards more popular items--there are a few popular items with thousands of ratings and the other majority of the items combined have fewer ratings than those few popular ones. Figure ~\ref{longtail} shows the long-tail distribution of item popularity in the famous MovieLens 1M dataset but the same distribution can be seen in other datasets as well.

Due to this imbalance property of the rating data, often algorithms inherit this bias and, in many cases, increase it by over-recommending the popular items and, therefore, giving them a higher opportunity of being rated by more users and, as a result, making them rated by more users and this goes on again and again--\textit{the rich get richer and the poor get poorer.} 

That by itself is very problematic and needs to be addressed using proper algorithms. However, what is often ignored is the fact that there are many users who are actually interested in non-popular items and they expect to get recommendations beyond some extremely popular items and blockbusters. What is worth noting is that users in a recommender system platform all should be considered as different stakeholders and their satisfaction should be taken care of by the algorithm \cite{abdollahpouri2019beyond}. An algorithm that is only capable of addressing the needs of a certain group of users (those whose interest is primarily concentrated on  popular items) will be in the risk of losing the interest of the rest of the users if their needs are not taken care of. 

In this paper we compare several recommendation algorithms and see how they perform in terms of keeping the ratio of popular and non-popular items according to what the users expect from the recommendations (i.e. the same ratio in their profile)

\subsection{User Propensity for Popular Items}
\begin{figure}
    \centering
    \includegraphics[height=1.7in]{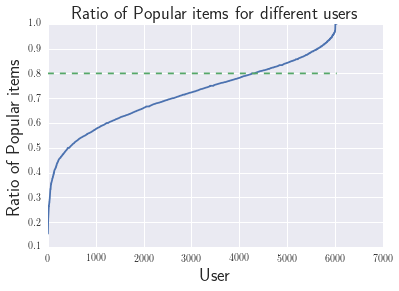}
    \caption{The ratio of popular items in users' profiles in MovieLens dataset}
    \label{fig:user_pop_ratio}
\end{figure}

As we mentioned earlier, there might be users who are interested in non-popular items and therefore the recommender algorithm should be able to address the needs of those users as well. In this section, we analyze the degree of interest that different types of users have towards popular items. Figure ~\ref{fig:user_pop_ratio} shows the ratio of popular items in the users' profiles.

The users have been sorted first based on the ratio of popular items in their profile then the data has been plotted. We can see that more than 4000 users are actually interested in at least 20\% non-popular items (from x-axis 0 to 4000). Even the other 2000 users also have some interest in non-popular items as the ratio of popular items, except for a dozen of users, is still between 80\% and 100\% meaning there is around 10-20\% non-popular items in their profile.

\subsection{Users Profile Size and Popularity Bias}

\begin{figure}
\centering
\SetFigLayout{3}{1}
  \subfigure[Correlation of profile size and the  number of popular items in user profile] {\includegraphics[width=2.4in]{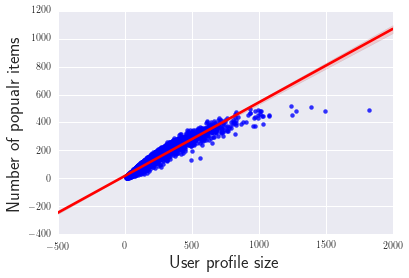}}
  \hfill
  \subfigure[Correlation of profile size and the  ratio of popular items in user profile]{\includegraphics[width=2.4in]{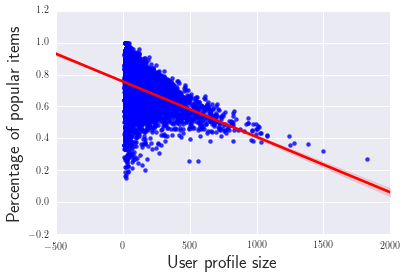}}
  \subfigure[Correlation of profile size and the average popularity of items in user profile]{\includegraphics[width=2.4in]{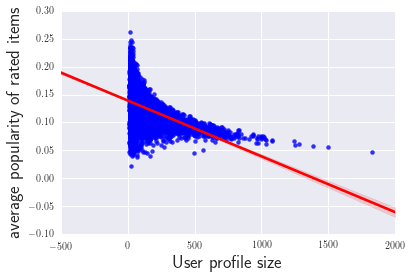}}
  \hfill
\caption{The correlation between the profile size and a) the number of popular items, b) the ratio of popular items and c) the average popularity of rated items in users' profiles}
\label{size_corr}
\end{figure} 
Before going any further, we want to see if a user's profile size has any correlation with the number of popular items she has in her profile. In this paper, we define an item to be popular if its popularity value falls within the top 20\% of item popularities.  Figure ~\ref{size_corr}-a shows a strong correlation between profile size and the number of popular items in users' profile which is kind of obvious due to the fact that having more items in a user's profile increases the likelihood of observing a popular item in it. So, we also show the correlation between the ratio of popular items in users' profiles and the profile size. Figure ~\ref{size_corr}-b shows this correlation is negative meaning the larger the profile size, the smaller the ratio of popular items is in a user's profile. This effect is to be expected. There are a small number of popular items. As a user's profile gets larger, they will eventually run out of popular items to rate and therefore must include more and more non-popular items. Figure ~\ref{size_corr}-c shows another perspective of popularity of the rated items in users' profile which is the average popularity of the rated items by each user. The popularity of each item is simply the ratio of users who have rated that item. Again in this figure we can see a negative correlation between profile size and the average popularity of rated items by a user indicating the more items a user has rated (larger profile size) the smaller the average popularity of her rated items is. 

These results provide interesting insights about users with lower propensity to rate popular items: these users also tend to interact with the system more and provide more ratings. Therefore, they tend to have a more profound effect on the performance of the recommender system and their contributions should not be ignored.

\section{Different Groups of Users in Terms of Propensity for Popularity}
In this section, we want to define three groups of users based on the ratio of popular items in their profile. That is, how interested they are in popular items. 

\begin{itemize}
    \item \textbf{Niche Users \textbf{(N)}}: These are the bottom 20\% of users in terms of the ratio of popular items in their profile. For these users, More than half of their profile consists of non-popular (longtail) items. 
    \item \textbf{Blockbuster-focused Users \textbf{(B)}}: These are the top 20\% of users in terms of the ratio of popular items in their profile. These users, on average, have more than 85\% popular items in their profile. 
    \item \textbf{Diverse Users \textbf{(D)}}: These are the users that are not \textit{Niche} neither are they \textit{Blockbuster-focused}. So the rest of the users fall within this category. 
\end{itemize}

\begin{figure}
\centering
\SetFigLayout{1}{2}
  \subfigure[The number of users in each group]{\includegraphics[width=3in]{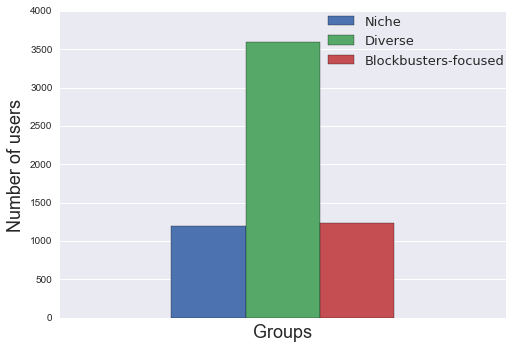}}
  \hfill
  \subfigure[The average profile size in different user groups]{\includegraphics[width=3in]{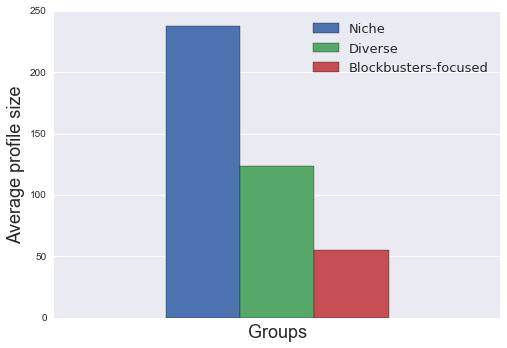}}
  \hfill
\caption{a) The number of users in three different groups N, D and B and b) The average profile size of the users in these groups}
\label{group_population}
\end{figure}

Figure ~\ref{group_population}-a shows the population of each of these different groups in MovieLens 1M dataset. As you can see the majority of the users are in group \textit{D} (i.e. diverse) with more than 3500 users. In addition, group \textit{N} and \textit{D} combined take up more than 4500 of the user population (around 75\% of the users).
 
Figure ~\ref{group_population}-b shows the average profile size of different user groups. Consistent with our previous analysis, niche users have larger profiles followed by diverse users. The Blockbuster-focused group has the shortest average profile size. 

\section{Algorithmic Propagation of Popularity Bias}

\begin{figure*}
\centering
\SetFigLayout{3}{2}
  \subfigure[Biased MF]{\includegraphics[width=2.2in]{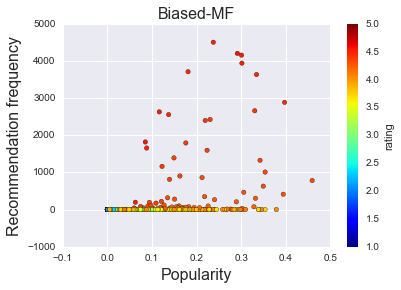}}
  \hfill
  \subfigure[SVD++]{\includegraphics[width=2.2in]{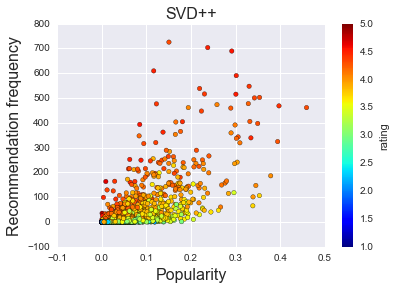}}
  \hfill
    \subfigure[Item KNN]{\includegraphics[width=2.2in]{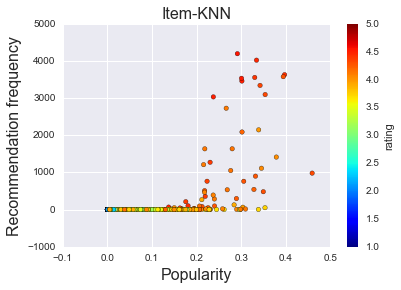}}
  \hfill
    \subfigure[User KNN]{\includegraphics[width=2.2in]{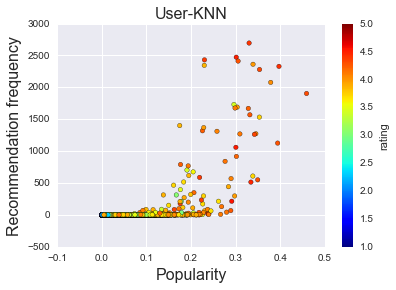}}
  \hfill
    \subfigure[Random]{\includegraphics[width=2.2in]{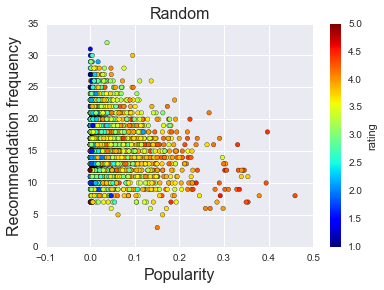}}
  \hfill
      \subfigure[Pop]{\includegraphics[width=2.2in]{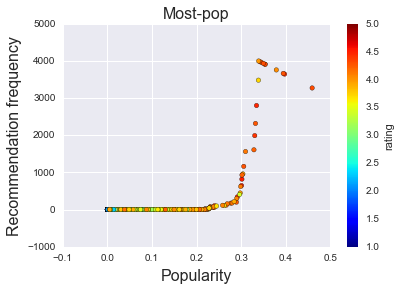}}
  \hfill
\caption{The correlation between the popularity of items and the number of times they are being recommended using different algorithms} \label{corr}
\end{figure*}

As we observed in previous section, there is a bias in the rating data: certain items are rated very frequently while many others are rated by fewer users. In this section, we show how different recommendation algorithms propagate this bias into their recommendations. We first look at their general performance without paying attention to how they perform for different users or groups of users. 
We tested several algorithms including \textit{User KNN}, \textit{Item KNN}, \textit{SVD++} and \textit{Biased Matrix factorization} for our experiments. For all these algorithms we set aside 80\% of the rating data as training set and the remaining 20\% as the test set. For each user in test set, a list of 10 items are recommended. Moreover, we tuned all these algorithms so they can achieve a similar precision (around 0.1) for a fair comparison of their popularity bias propagation. In addition, we also used two simple \textit{Most popular} and \textit{Random} algorithms.
Figure ~\ref{corr} shows the correlation between the number of times an item is rated and the number of times it is recommended by different algorithms. 

It can be seen that in all algorithms except for the \textit{random} there are many items that are almost never recommended (the items fall on the horizontal tail of the scatter plot). In addition, the \textit{Most popular} algorithm seems to have the strongest correlation between the number of times an item is being rated and the number of times it is recommended, as it was expected, followed by \textit{User KNN} and \textit{Item KNN} which also have a strong correlation. Looking at the plot for the \textit{Biased MF} algorithm, it seems there is not a significant correlation between how often an item is rated and how often it is recommended. Nevertheless, its very sparse scatter plot shows the number of recommended items is still low. In this algorithm there are a lot of items that are recommended very rarely (even some of the ones that are rated more frequently) while some few items are being recommended more frequently. \textit{SVD++} also shows a positive correlation between the number of times an item is rated and is recommened. The random algorithm obviously does not show any correlation between the number of times an item is rated and the number of times it is recommended. For each point on the scatter plot, you can also see the average rating for the corresponding item to illustrate the quality of these recommended items using each algorithm. 
 
\subsection{Popularity Bias of Recommendations in Different User Groups}
In this part, we want to look at how these different algorithms perform in terms of keeping the right ratio of popular and non-popular items recommended to different user groups according to their expected value of such ratio. 

\begin{figure*}
\centering
\SetFigLayout{3}{2}
  \subfigure[Biased MF]{\includegraphics[width=2in]{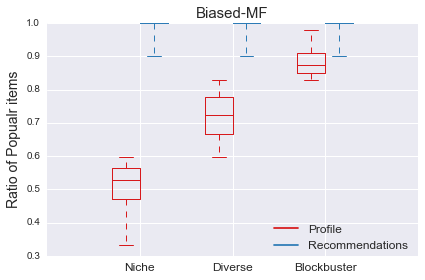}}
  \hfill
  \subfigure[SVD++]{\includegraphics[width=2in]{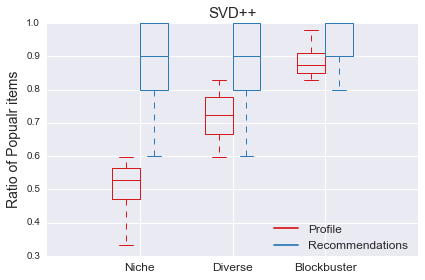}}
  \hfill
    \subfigure[Item KNN]{\includegraphics[width=2in]{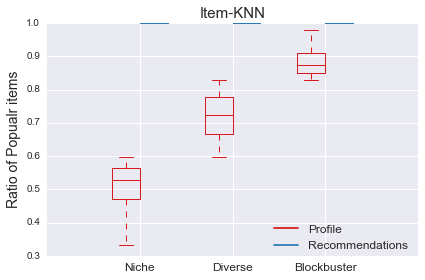}}
  \hfill
    \subfigure[User KNN]{\includegraphics[width=2in]{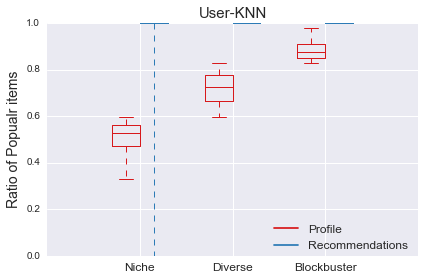}}
  \hfill
    \subfigure[Random]{\includegraphics[width=2in]{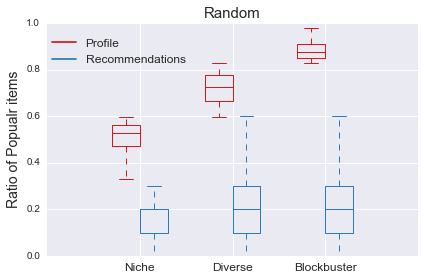}}
  \hfill
      \subfigure[Pop]{\includegraphics[width=2in]{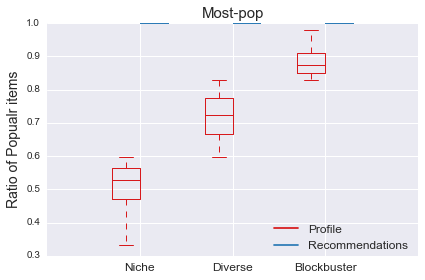}}
  \hfill
\caption{The ratio of popular items in the profiles of users in different groups versus the ratio of popular items in their recommendations } \label{boxplot}
\end{figure*}

Figure ~\ref{boxplot} shows the ratio of popular items in the profiles of users in different groups versus the ratio of popular items in their recommendations. As you can see, for all algorithms except for random, the  ratio of popular items in the recommendations is significantly increased for all user groups confirming how these algorithms impose popularity on users. Specifically, the Niche users are seeing the largest discrepancy of such ratio imposed by the algorithms. However, the recommendations generated by SVD++ seem to be more consistent with what the users expect to get in terms of the right ratio of popular and non-popular items.

In addition, we define the \textit{Group Average Popularity (GAP (g))} metric that measures the average popularity of items in the profiles of users in a certain group \textit{g} or their recommendation lists. If we look at the recommendations given to a group of users then \textit{GAP} measures the average popularity of the recommended items to the users in that group. On the other hand, if we look at the users' profiles, then \textit{GAP} measures the average popularity of the items rated by the users in that group. 

\begin{equation}
    GAP(g)=\frac{\sum_{u \in g}\frac{\sum_{i \in p_u} \phi(i)}{|p_u|}}{|g|}
\end{equation}

were \textit{g} is the group of users (in our case it is either \textit{N}, \textit{D} or \textit{B}). $\phi(.)$ is the popularity of a certain item (i.e. the number of times it is rated divided by the total number of users)  and $p_u$ is the list of items in the profile of user \textit{u}. 
We represent the GAP value in user profiles as $GAP_p$ and the GAP for recommendations by $GAP_r$. 
For each recommendation algorithm, We measure the change in \textit{GAP} which is the amount of unwanted popularity in the recommendations imposed by the algorithms to each group. The value of $\Delta GAP=0$ means a fair representation of users' preferences towards item popularity in the recommendations. 

\begin{equation}
    \Delta GAP(g)=\frac{GAP(g)_r-GAP(g)_p}{GAP(g)_p}
\end{equation}

\begin{figure}[t]
    \centering
    \includegraphics[width=3.3in]{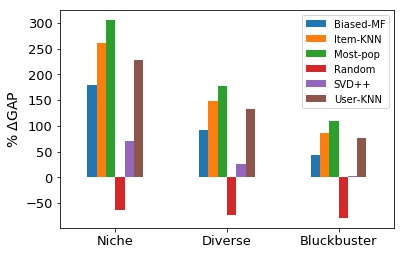}
    \caption{The $\Delta$ GAP of different algorithms in user groups}
    \label{fig:GAP}
\end{figure}

Figure ~\ref{fig:GAP} shows the change in the group average popularity ($\Delta GAP$) in different groups using different algorithms. \textit{Most popular} algorithm imposes the highest positive change in GAP followed by item KNN. Random has negative $\Delta$GAP meaning the recommendations on average are less popular compared to the average popularity of users profiles in each group. We can also see that the niche users have the largest $\Delta GAP$ for all algorithms. 

 These results show the unfair nature of popularity bias in the recommendations and how its effect  on different users varies based on how interested the users are in popular items.
\section{Discussion}
In this section we would like to summarize the answers we found to the research questions we listed in section 1. 
\begin{itemize}
     \item \textbf{Answer to RQ1: }  Our analysis showed that indeed not everyone has the same level of interest in popular items. We found that there are large number of users who are also interested in non-popular items and expecting to get some of those as recommendations. We showed that users with larger profiles are typically less likely to have exclusively rated popular items and they do have some interest in long-tail or non-popular items.
     
     \item \textbf{Answer to RQ2: } Our results showed that the recommendations generated by all the algorithms were extremely unfair to users with lesser interest in popular items as almost all the recommendations they received were popular items. That showed not every user is affected by the popularity bias the same as other users.   
 
\end{itemize}
\section{Conclusion and Future Work}
Due to the popularity bias in rating data, many recommendation algorithms propagate this bias even further by recommending popular items more frequently while not giving enough exposure to less popular ones. In this paper, we investigated the popularity bias from a different point of view: the user's perspective. We showed that even though there are many users who are interested in getting some non-popular items, they almost do any get any of those in their recommendations. We further defined three groups of users (niche, diverse and Blockbuster-focused) based on the level of their interest towards popular items. We observed that all algorithms recommend items that are much more popular than what the users in those groups have rated. Especially the niche users get really inappropriate recommendations when the level of popularity of the recommendations is compared to what they are interested in. For future work, we will use more datasets for our analysis. We also want to investigate other measures of satisfaction for different user groups such as how relevant their recommendations are, how diverse they are etc.

\bibliographystyle{ACM-Reference-Format}
\bibliography{main.bib}

\end{document}